\documentclass[%
 preprint,
singleocolumn,
 amsmath,amssymb,
 aip,
 jmp,
]{revtex4-1}

\usepackage{textcomp}
\usepackage{graphicx}% Include figure files
\usepackage{dcolumn}% Align table columns on decimal point
\usepackage{bm}% bold math
\begin{document}

%\preprint{APS/123-QED}

\title{Charge-voltage relation for a universal capacitor}
%\thanks{A footnote to the article title}%

\author{Vikash Pandey}
\email{vikash.pandey@ddn.upes.ac.in}

\affiliation{%
Dept.~of Mathematics, University of Petroleum and Energy Studies (UPES), Dehradun, Uttarakhand, PIN: 248007, INDIA
%\textbackslash
}

%\affiliation{%
%University of Oslo, P.O. Box 1066, NO-0316, Oslo, NORWAY
% \footnote[1]{Affiliation when most of the work was completed.}
%\textbackslash
%}%

\date{\today}% It is always \today, today,
             %  but any date may be explicitly specified

\begin{abstract}

Most capacitors do not satisfy the conventional assumption of a constant capacitance. They exhibit memory which is often described by a time-varying capacitance. It is shown that the classical relation, $Q\left(t\right)=CV\left(t\right)$, that relates the charge, $Q$, with the capacitance, $C$, and the voltage, $V$, is not applicable for capacitors with a time-varying capacitance. The expression for the current, $dQ/dt$, that is subsequently obtained following the substitution of $C$ by  $C\left(t\right)$ in the classical relation corresponds to an inconsistent circuit. In order to address the inconsistency, I propose a charge-voltage relation according to which the charge on a capacitor is expressed by the convolution of its time-varying capacitance with the first-order time-derivative of the applied voltage, i.e., $Q\left(t\right)=C\left(t\right)\ast dV/dt$. This relation corresponds to the universal capacitor which is also known as the fractional capacitor among the fractional calculus community. Since the fractional capacitor has an inherent connection with the universal dielectric response that is expressed by the century old Curie-von Schweidler law, the finding extends to the study of dielectrics as well.
 
\end{abstract}

%\keywords{Suggested keywords}%Use showkeys class option if keyword
                              %display desired
\maketitle

%\tableofcontents

---------------

It is well established that almost all capacitors exhibit memory because of the inherent time-dependent relaxation of the dielectric media that is sandwiched in between their parallel plates \cite{Jonscher1977,Jonscher1977a,Westerlund1991,Westerlund1994,Ershov1997,Uchaikin2009}. An efficient way to represent a capacitor's memory is to assume a time-varying capacitance, $C\left(t\right)$.  Such an assumption has been used in the study of, solid state devices \cite{Lee1993,Brauer1998}, time-varying storage components \cite{Biolek2007,Richards2012}, energy accumulation \cite{Mirmoosa2019}, brain microvascularity \cite{Jo2015}, and biomimetic membranes \cite{Najem2019}. But most of these  cited references use the equation, 
\begin{equation}
Q\left(t\right)=C\left(t\right)V\left(t\right),\label{eq:charge_relation}
\end{equation}
for their investigations. The equation is directly motivated from the classical charge-voltage relation of a capacitor, $Q=CV$, where, $Q$ is the accumulated charge, $C$ is the constant capacitance, and $V$ is the applied voltage, and $t$ is the time. The current is then obtained from \eqref{eq:charge_relation}  following the product rule of differentiation as:
\begin{equation}
I\left(t\right)=\dot{Q}\left(t\right)=C\left(t\right)\dot{V}\left(t\right)+V\left(t\right)\dot{C}\left(t\right),\label{eq:current_relation}
\end{equation}
where the number of over-dots represent the order of differentiation
with respect to time.
Here, it is necessary to emphasize that even though the classical relation is only applicable for capacitors with a constant capacitance   \cite{Westerlund1994}, \eqref{eq:charge_relation} and \eqref{eq:current_relation} have been widely used by physicists and engineers for describing capacitors with time-varying capacitances. Unfortunately, this has been overlooked in the time-varying circuit theory \cite{Richards2012}, as well as in the basic circuit theory \cite{Desoer2010}. Even further the circuit simulation tools such as those from Matlab and Micro-Cap \cite{Biolek2007} also use \eqref{eq:current_relation} to model current through capacitors with time-varying capacitances. I will later show that \eqref{eq:current_relation} seems to be incorrect for time-varying capacitors which makes those results doubtful that were led by that ignorance. 

The industrial manufacturers of capacitors, in lack of a better model than that expressed by \eqref{eq:charge_relation}, circumvent the lacking by defining capacitance at $1$ kHz \cite{Westerlund1991}. Interestingly, an attempt was made by Westerlund to resolve this problem by proposing a charge-voltage relation for a universal capacitor model \cite{Westerlund1991}. But contrary to the expectation that work did not attracted significant interest among the electrical engineers community because of three reasons. First, the lack of a closed-form expression for the charge-voltage relation, see the abstract and equations (16) and (17) in \cite{Westerlund1991}. Second, though the results were motivated from the experimental observations, the underlying issue with the classical relation was neither discussed nor proved analytically. Third, the acceptability of the fractional derivatives among the scientific community was relatively less three decades ago than it is now \cite{Podlubny2002,Machado2015}. This letter aims to present a closed-form expression for the charge-voltage relation of a universal capacitor model that also addresses the limitations of the classical model. However, I first show the inconsistency that arises from the classical equations, \eqref{eq:charge_relation} and \eqref{eq:current_relation}, through a simple circuit analysis. 

\begin{figure}[h!]
\begin{centering}
\includegraphics[width=1\columnwidth]{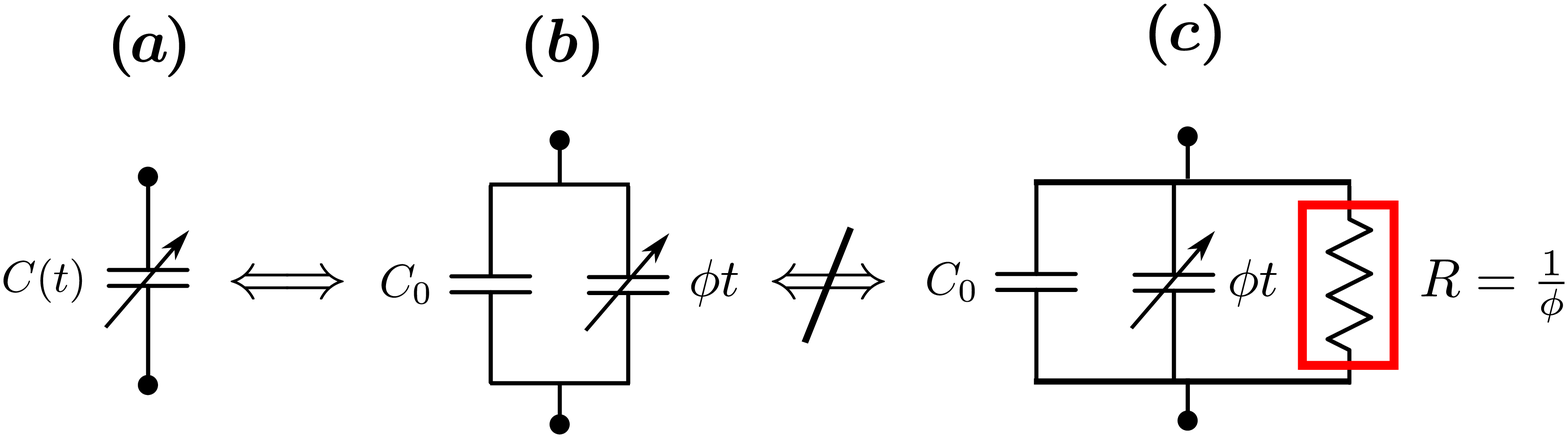}
\par\end{centering}
\caption{(a) Symbol for a time-varying capacitor with a time-varying capacitance, $C\left(t\right)$. (b) Equivalent model of (a), assuming a linearly time-varying capacitance, $C\left(t\right)=C_{0}+\phi t$. (c) The inequivalent model with an undesired resistor that emerges from the application of the classical relation,  $Q\left(t\right)=C\left(t\right)V\left(t\right)$, to (b).}
\begin{centering}
\label{Time_capacitor}
\par\end{centering}
\end{figure}

Let there be a capacitor with a time-varying capacitance, $C\left(t\right)$. The capacitance is a sum of
the constant geometric capacitance, $C_{0}$, and a time-varying capacitance, $C_{\phi}\left(t\right)$. The time-varying part of the capacitance is due to the dielectric media present in the capacitor, see equation (2) in \cite{Westerlund1991}. Further assuming a linearly time-varying capacitance, $C_{\phi}\left(t\right)=\phi t$, such that, $\phi$, is a positive real constant, I have,
\begin{equation}
C\left(t\right)=C_{0}+C_{\phi}\left(t\right)=C_{0}+\phi t.\label{eq:time_var_cap}
\end{equation}
Since capacitances add in parallel circuits, the equivalent circuits are shown in Figs.~\ref{Time_capacitor}(a) and (b).
On substituting \eqref{eq:time_var_cap} in \eqref{eq:charge_relation} and \eqref{eq:current_relation}, I have the following,
\begin{equation}
Q\left(t\right)=\left(C_{0}+\phi t\right)V\left(t\right)\text{, and }\label{eq:curie_charge}
\end{equation}
\begin{equation}
I\left(t\right)=C_{0}\dot{V}+\phi t\dot{V}+V\phi.\label{eq:curie_current}
\end{equation}
On carefully observing the three additive terms on the right hand side of \eqref{eq:curie_current}, I find that the first
two corresponds to capacitor currents that flow through the capacitors of capacitances, $C_{0}$ and $C_{\phi}$, respectively. But the third term is the Ohmic current that flows through a resistor of equivalent resistance, $R=1/\phi$. Since currents add in parallel branches, \eqref{eq:curie_current}, corresponds to a parallel combination of the three elements as shown in Fig.~\ref{Time_capacitor}(c). Clearly, this is not equivalent to the Fig.~\ref{Time_capacitor}(b). This anomaly may also be verified as follows. On imposing the initial condition at time, $t=0$, I have from \eqref{eq:curie_current}, the current, $I_{0}=C_{0}\dot{V}+V\phi$, instead of the expected, $I_{0}=C_{0}\dot{V}$. The root cause of this problem can be traced back to the classical relation, \eqref{eq:charge_relation}, which is not valid for a time-varying capacitance \cite{Westerlund1991,Westerlund1994}. The underlying reason behind the inequivalence of Figs.~\ref{Time_capacitor}(b) and  \ref{Time_capacitor}(c) is that the traditional charge-voltage relation assumes a linearly time-invariant system, i.e., $Q\left(t\right)=f\left(V\left(t\right)\right)$. In contrast, a time-varying capacitance invokes a time-variant system, $Q\left(t\right)=f\left(V\left(t\right), t\right)$, i.e., the capacitor remembers the applied voltage that it was subjected to, in the past \cite{Westerlund1994}. The classical relation  leads to a term-by-term multiplication of $C\left(t\right)$ and $V\left(t\right)$ at any given instant of time, $t$, and therefore it does not take the capacitor memory into account.

The memory characteristic of the capacitor is taken into account through the following proposed charge-voltage relation,
\begin{equation}
Q\left(t\right)=C\left(t\right)\ast\dot{V}\left(t\right),\label{eq:new_charge}
\end{equation}
where, ``*" represents the convolution operation. The relation is different than (5) from \cite{Fouda2020} which seems to have a dimensional inconsistency. Further, substituting \eqref{eq:time_var_cap} in \eqref{eq:new_charge} and following the derivative property of the convolution, the capacitor current is then obtained as:
\begin{equation}
I\left(t\right)=C\left(t\right)\ast\ddot{V}\left(t\right)=\left[C_{0}\ast\ddot{V}\left(t\right)\right]+\left[\phi t\ast\ddot{V}\left(t\right)\right].\label{eq:convolv_current}
\end{equation}
Interestingly, convolutions are also common in the field of fractional derivatives that provide a robust mathematical framework to study time-variant systems. The fractional framework has proven its versatility in describing systems that exhibit both spatial memory \cite{Pandey2016} and temporal memory \cite{Pandey2016a}. According to Caputo \cite{Mainardi2010}, the fractional derivative for a continuous, causal function, $f\left(t\right)$, is defined as the convolution of a regular integer-order derivative with a power-law memory kernel, $\Phi_{\alpha}\left(t\right)$, as:
\begin{equation}
\frac{d^{\alpha}}{dt^{\alpha}}f\left(t\right)\triangleq\dot{f}\left(t\right)\ast\Phi_{\alpha}\left(t\right),\text{ }\Phi_{\alpha}\left(t\right)=\frac{t^{-\alpha}}{\Gamma\left(1-\alpha\right)},\text{ }0<\alpha<1,\label{eq:frac_deriv}
\end{equation}
where, $\alpha$ is the order, and $\Gamma\left(\cdot\right)$ is the Euler Gamma function. For negative values of $\alpha$, \eqref{eq:frac_deriv} corresponds to a fractional integral. The Fourier transform property, $\mathcal{F}\left[d^{\alpha}f\left(t\right)/dt^{\alpha}\right]=\left(i\omega\right)^{\alpha}\boldsymbol{f}\left(\omega\right)$, of fractional derivatives, where $\omega$ is the angular frequency, confirm that they are a mere generalization of the regular integer-order derivatives. Furthermore, in appropriate limiting conditions, fractional derivatives asymptotically converge to the integer-order derivatives. In recent years, a connection between the fractional derivatives and the physics of complex media has also been established \cite{Pandey2016a,Pandey2016b,Pandey2016c,Pandey2020a}. It is worth noticing that the expressions for, $Q$ and  $I$, in \eqref{eq:new_charge} and \eqref{eq:convolv_current}, are actually motivated from the fractional derivatives. The expression for the current, \eqref{eq:convolv_current},  when seen in light of \eqref{eq:frac_deriv}, gives, 
\begin{equation}
I\left(t\right)=C_{0}\dot{V}\left(t\right)+\phi V\left(t\right).\label{eq:new_current}
\end{equation}
Also, at any instant of time, $V\left(t\right)\equiv t \dot{V}\left(t\right)$, which when substituted back in \eqref{eq:new_current}, leads to,
\begin{equation}
    \begin{split}
        I\left(t\right)=I_{C_{0}}\left(t\right)+I_{C_{\phi}}\left(t\right),\\
        \text{ where }  I_{C_{0}}\left(t\right) = C_{0}\dot{V}\left(t\right), \text{ and }  I_{C_{\phi}} = C_{\phi} \dot{V}\left(t\right),
    \end{split}
    \label{eq:super_new_current}
\end{equation}
are the capacitor currents that flow through the capacitors of capacitances, $C_{0}$ and $C_{\phi}$, respectively. It can be seen that \eqref{eq:super_new_current} is equivalent to the current flowing in the circuit shown in Fig.~\ref{Time_capacitor}(b). Thus the inequivalence that arose due to the conventional charge-voltage relation, \eqref{eq:charge_relation}, is resolved through the convolution relation, \eqref{eq:new_charge}. It can be inferred that the additional unwanted term, $V\phi$, in \eqref{eq:curie_current}, that had its origin from the term, $V\left(t\right)\dot{C}\left(t\right)$, from \eqref{eq:current_relation}, has vanished. Therefore, if the last term of \eqref{eq:current_relation} is neglected and a direct substitution of, $C\left(t\right)$, from  $\left(\ref{eq:time_var_cap}\right)$, is made in the first term of \eqref{eq:current_relation}, then I get the same result as that from \eqref{eq:super_new_current}. So, it can be concluded that the last term, $V\left(t\right)\dot{C}\left(t\right)$, in  \eqref{eq:current_relation}, is not required at all. Fortunately, this has been experimentally verified as well \cite{Jadli2020}. It is also possible to obtain  \eqref{eq:super_new_current} from \eqref{eq:convolv_current} using the standard convolution integral. However if the time-varying capacitance is expressed in the form of a power-law, then fractional framework turns out to be a readily available tool for their analysis. 
It should be emphasized that if the condition, $V\left(t\right)\equiv t \dot{V}\left(t\right)$, was applied on the last term of \eqref{eq:curie_current}, that would have led to a capacitor current identical to the second term of the same equation. The resulting circuit would then be again inequivalent to the circuit shown in Fig.~\ref{Time_capacitor}(b). This further stresses on the inapplicability of the classical charge-voltage relation.

Further, even though I have assumed a linearly time-varying capacitance, the proof that I have presented here can be generalized to all power-law forms of the time-varying capacitance using fractional derivatives. On replacing, $C\left(t\right)$, from  \eqref{eq:convolv_current}, by, $C_{0}\left(\tau/t\right)^{\alpha-1}/\Gamma\left(2-\alpha\right)$, and then interpreting the resulting equation in light of \eqref{eq:frac_deriv}, the expression for the current through a fractional capacitor is obtained as \cite{Elwakil2010,Fouda2020}: 
\begin{equation}
I\left(t\right)=C_{0}\tau^{\alpha-1}\left[\frac{t^{1-\alpha}}{\Gamma\left(2-\alpha\right)}\ast\ddot{V}\left(t\right)\right]=C_{f}\frac{d^{\alpha}}{dt^{\alpha}}V\left(t\right),\label{eq:last_frac_current}
\end{equation}
where, $C_{f}=C_{0}\tau^{\alpha-1}$ is the pseudocapacitance and $\tau$ is the characteristic time constant. Therefore, it is inferred that \eqref{eq:new_charge} corresponds to the charge-voltage relation for a fractional-capacitor. The fractional capacitor has an interpolating behavior between a resistor and a capacitor for, $0<\alpha<1$. Besides, because of its constant phase angle, $|\alpha \pi/2|$, property, the fractional capacitor is also referred as the \textit{constant phase element}. Fractional capacitors for arbitrary values of $\alpha$ are fabricated in laboratories \cite{Elshurafa2013,Tsirimokou2016,John2017a}, and have applications in the modeling of biological media \cite{Jo2015,Najem2019}, dielectric media\cite{Jonscher2001,Luo2004,XU2004,Jameson2006,Ning2008}, supercapacitors \cite{Allagui2018a,Allagui2018b}, and electrochemical capacitors \cite{Martynyuk2018}.

Though it seems that $\phi$ can be easily extracted from the slope of $C$ versus $t$ using \eqref{eq:time_var_cap}, it is the current that is measured in experiments. But the convoluted form of the current, \eqref{eq:last_frac_current}, makes it difficult. However, an estimation of $\phi$ is still possible from the fact that \eqref{eq:last_frac_current} is an equivalent representation of the century old Curie-von Schweidler law \cite{Jonscher1977a,Westerlund1991,Westerlund1994}. The law has a characteristic power law form, $I\left(t\right)\propto t^{-\alpha}$, and because of its universal applicability, \eqref{eq:last_frac_current}, is regarded as the expression for current of a universal capacitor \cite{Jonscher1977,Westerlund1994}. Interestingly, \eqref{eq:new_charge} was used as an intermediate step in the derivation of the Curie-von Schweidler law from physical principles that also gave a physical interpretation of the law for the first time \cite{Pandey2020a}. Accordingly from \cite{Pandey2020a},  I have, $\alpha = 1/\left(\boldsymbol{R}\phi\right)$, and $\tau=C_{0}/\phi$, where $\boldsymbol{R}$ represents the inherent resistance of a dielectric media. In the case of the circuit modelling of dielectrics, the resistance is represented by a resistor that is connected in series with a time-varying capacitance, see Fig.~1(a) in \cite{Pandey2020a}. Such a circuit has been experimentally studied in \cite{Uchaikin2009} with the parameter values, $\alpha=0.998$, $\boldsymbol{R}=200 \text{ k}\Omega$, and $C_{0}=2 \text{ }\mu\text{F}$. Then using \eqref{eq:new_charge} and results from \cite{Pandey2020a}, $\phi=1/(\boldsymbol{R}\alpha)= 5.01 \text{ }\mu\text{F}/s$, from which, the time constant is calculated as, $\tau=C_{0}/\phi\approx0.3992 \text{ }s$. This is very close to the expected value of, $\tau=\boldsymbol{R}C_{0}=0.4\text{ }s$. This further consolidates the results presented in this manuscript.

If the capacitance is assumed to be a constant, i.e., $C_{\phi}=0$, then results from \eqref{eq:new_charge} and \eqref{eq:convolv_current},  reduce to the classical relations, $Q\left(t\right)=C_{0}V\left(t\right)$ and $I\left(t\right)=C_{0}\dot{V}\left(t\right)$, respectively, which is expected for a time-invariant system. This can be witnessed from the first term that appears on the right hand side of \eqref{eq:super_new_current}. The same also mirrors from \eqref{eq:last_frac_current}, for $\alpha=1$. Therefore, the convolution relations expressed by \eqref{eq:new_charge} and \eqref{eq:convolv_current}, which also correspond to the fractional capacitor, should be seen as relations that complete the bigger picture and yet retain the beauty of the classical relations. Moreover, since the fractional capacitor is also considered as the \textit{universal capacitor}, the relation, $Q\left(t\right)=C\left(t\right)\ast\dot{V}\left(t\right)$, may be regarded as the \textit{universal} charge-voltage relation for capacitors. I believe this finding may further boost the emerging field of fractional-order circuits and systems.

\section*{Acknowledgment}

The author would like to thank his former colleagues, Asst.~Profs.~Jaina Mehta and Maryam Kaveshgar, from Ahmedabad University, Gujarat, for going through the manuscript and for the fruitful discussions that the author had with them.

\end{document}